\documentclass[a4paper,11pt]{article}
\usepackage{pos}
\usepackage{braket}
\usepackage{bm}
\usepackage{slashed}
\usepackage{cleveref}
\usepackage{mciteplus}

\def\p{{\boldsymbol p}}

\def\r{{\boldsymbol r}}
\def\R{{\boldsymbol R}}

\def\l{{\boldsymbol l}}
\def\v{{\boldsymbol v}}

\DeclareMathOperator{\Tr}{Tr}
\newcommand*\diff{\mathop{}\!\mathrm{d}}
\newcommand{\GeV}{{{\,}\textrm{GeV}}}
\newcommand{\expconfig}[1]{{\langle #1 \rangle_m}}

\title{On the momentum broadening of in-medium jet evolution using a light-front Hamiltonian approach}
\ShortTitle{In-Medium Jet Evolution: Light-Front Hamiltonian Approach}

\author*[a]{Meijian Li}
\author[b,c]{Tuomas Lappi}
\author[d,e]{Xingbo Zhao}
\author[a]{Carlos A. Salgado}

\affiliation[a]{Instituto Galego de Fisica de Altas Enerxias (IGFAE), Universidade de Santiago de Compostela, E-15782 Galicia, Spain}
\affiliation[b]{Department of Physics, P.O. Box 35, FI-40014 University of Jyv\"{a}skyl\"{a},
Finland}
\affiliation[c]{
Helsinki Institute of Physics, P.O. Box 64, FI-00014 University of Helsinki,
Finland
}
\affiliation[d]{Institute of Modern Physics, Chinese Academy of Sciences, Lanzhou 730000, China}
\affiliation[e]{University of Chinese Academy of Sciences, Beijing 100049, China}

\emailAdd{meijian.li@usc.es}
\emailAdd{tuomas.v.v.lappi@jyu.fi}
\emailAdd{xbzhao@impcas.ac.cn}
\emailAdd{carlos.salgado@usc.es}

\abstract{
  We have developed a non-perturbative light-front Hamiltonian formalism to simulate the real-time evolution of a quark state in a SU(3) colored medium, with a series of works. 
  In this proceeding article, we focus on the transverse momentum broadening of an in-medium quark jet.
  We perform the numerical simulation of the quark jet evolution in the $\ket{q}+\ket{qg}$ Fock space at various medium densities.
  By analyzing the resulting jet light-front wavefunction, we extract the gluon emission rate and the non-eikonal quenching parameter. 
  Additionally, we provide the analytical derivation of the eikonal expectation value of the quark-gluon state's transverse momentum for any color configuration and arbitrary spatial distribution.
  This study can help understand jet momentum broadening beyond the eikonal limit.
}

\FullConference{HardProbes2023\\
 26-31 March 2023\\
 Aschaffenburg, Germany\\}

\begin{document}
\maketitle

\section{Introduction}
In heavy-ion collisions, energetic quarks and gluons are produced at early stages, propagating through the dense and hot medium. Similar processes happen in deeply inelastic scattering where quark and gluon jets traverse cold nuclear matter.
We developed a non-perturbative computational method, the time-dependent Basis Light-Front Quantization (tBLFQ)~\cite{1stBLFQ}, for simulating the evolution of a quark jet inside a classical color background field, first for a $\ket{q}$ state~\cite{Li:2020uhl}, then also including $\ket{qg}$ components~\cite{Li:2021zaw, Li:2023jeh}. 
Unlike pQCD-based approaches, tBLFQ calculates the evolution process at the amplitude level, and enables relaxation of the eikonal and collinear radiation approximations.

\section{Methodology}\label{sec:method}
We consider a high-energy quark jet moving in the positive $z$ direction, traversing a medium moving in the negative $z$ direction. 
We treat the quark as a quantum state and the medium as an external background field, with the interaction occurring over a finite distance $0\le x^+\le L_\eta$. 

\subsection{The light-front Hamiltonian in the $\ket{q}+\ket{qg}$ space}\label{sec:LFH}
The Lagrangian for the process being considered is the QCD Lagrangian with an external field,
\begin{align}\label{eq:Lagrangian}
 \mathcal{L}=-\frac{1}{4}{F^{\mu\nu}}_a F^a_{\mu\nu}+\overline{\Psi}(i\gamma^\mu  D_\mu -  m_q)\Psi\;,
\end{align}
where $F^{\mu\nu}_a\equiv\partial^\mu C^\nu_a-\partial^\nu C^\mu_a-g f^{abc}C^\mu_b C^\nu_c$, $D^\mu\equiv \partial^\mu +ig C^\mu$, and $ C^\mu= A^\mu + \mathcal{A}^\mu$ is the sum of the quantum gauge field $ A^\mu$ and the background gluon field $\mathcal{A}^\mu$.  
The light-front Hamiltonian is obtained through Legendre transformation in the light-cone gauge $A^+=\mathcal A^+ = 0$. 
In the truncated Fock space $\ket{q}+\ket{qg}$, it consists of three parts, $P^-(x^+)=P_{KE}^- + V_{qg}+V_{\mathcal{A}}(x^+)$, the kinetic energy, the interaction between the quark and the dynamical gluon, and the medium interaction,
\begin{subequations}
  \begin{align}
      P_{KE}^-=&\int\diff x^-\diff^2 x_\perp
      \bigg\{
      -\frac{1}{2}A^j_a{(i\nabla)}^2_\perp A_j^a
      +\frac{1}{2}\overline{\Psi}\gamma^+\frac{m_q^2-\nabla_\perp^2}{2i\partial_-}\Psi
      \bigg\}
      \;,\\
      V_{q g}=&\int \diff x^- \diff^2 x_\perp
       g\bar{\Psi}\gamma^\mu T^a\Psi A^a_\mu
      \;,\\
        V_{\mathcal{A}}(x^+)=&
        \int \diff x^- \diff^2 x_\perp
         g\overline{\Psi}\gamma^+ T^a\Psi\mathcal{A}^a_+(x^+)+
         g f^{abc} \partial^+ A^i_b A^c_i \mathcal{A}^a_+(x^+)
         \;.    
   \end{align}
\end{subequations}

The background field $\mathcal{A}^\mu$ accounts for the target, and we describe it using the McLerran-Venugopalan (MV) model~\cite{McLerran:1993ni}. It is a classical field satisfying the reduced Yang-Mills equation,
\begin{align}\label{eq:poisson}
 (m_g^2-\nabla^2_\perp )  \mathcal{A}^-_a(\vec{x}_\perp,x^+)=\rho_a(\vec{x}_\perp,x^+)\,,\quad
 \braket{\rho_a(x)\rho_b(y)}=g^2\tilde\mu^2\delta_{ab}\delta^2(\vec{x}_\perp-\vec{y}_\perp)\delta(x^+-y^+)\;.
\end{align}
The saturation scale is related to $\tilde\mu$ as
$Q_s^2=C_F (g^2\tilde\mu)^2L_\eta/(2\pi)
 $, with $C_F=(N_c^2-1)/(2N_c)=4/3$.

\subsection{Evolution of the state in a basis representation}
The evolution of quantum states is governed by the time-evolution equation on the light front. 
In the interaction picture (denoted by the subscript $I$), the equation reads
\begin{align}\label{eq:ShrodingerEq}
 i\frac{\partial}{\partial x^+}\ket{\psi;x^+}_I=\frac{1}{2}V_I(x^+)\ket{\psi;x^+}_I\;,
\end{align}
with the interaction Hamiltonian $V_I(x^+)=e^{i\frac{1}{2}P^-_{KE}x^+}V(x^+)e^{-i\frac{1}{2}P^-_{KE}x^+}$. The interaction picture state is related to the Schrödinger picture state by $\ket{\psi;x^+}_I=e^{i\frac{1}{2}P^-_{KE}x^+}\ket{\psi;x^+}$.

We implement a non-perturbative treatment by decomposing the time-evolution operator into many small steps of the light-front time $x^+$, then solving each timestep in the sequence numerically,
\begin{align}\label{eq:time_evolution_exp}
 \begin{split}
  \ket{\psi;x^+}_I=\mathcal{T}_+ &e^{-\frac{i}{2}\int_0^{x^+}\diff z^+V_I(z^+)}\ket{\psi;0}_I
   =\lim_{n\to\infty}\prod^n_{k=1}\mathcal{T}_+ e^{-\frac{i}{2}\int_{x_{k-1}^+}^{x_k^+}\diff z^+V_I(z^+)}\ket{\psi;0}_I
 \;,
 \end{split}
\end{align}
with $\mathcal{T}_+$ denoting light-front time ordering. 
The step size is $\delta x^+ \equiv x^+/n$, and the intermediate time is $x_k^+=k\delta x^+ (k=0,1,2,\ldots,n)$ with $x_0^+=0$ and $x_n^+=x^+$.

We use a lattice with periodic boundary conditions in the transverse dimensions $\vec x_\perp$, ranging in $[-L_\perp, L_\perp]$ with $2N_\perp$ sites such that $a_\perp=L_\perp/N_\perp$ is the lattice spacing, and a loop with (anti-)periodic boundary condition in the $x^-$ direction, of length $2L$, for the gluon(quark).
The lattice introduces infrared~(IR) and ultraviolet~(UV) cutoffs in the transverse momentum space $\vec p_\perp$, $\lambda_{IR}=d_p=\pi/L_\perp$ and $\lambda_{UV}=\pi/a_\perp$.
The longitudinal momentum $p^+$ is quantized in units of $2\pi/L$, and the gluon(quark) is allowed to take a positive (half-)integer number in this unit. The total momentum is denoted as $p^+= K 2\pi/L$ with $K$ a half-integer. 
Then the longitudinal momentum fraction of the gluon, $z\equiv p^+_g/p^+$, has a resolution of $1/K$.

In the chosen discrete basis representation, the state is a column vector of basis coefficients, and the Hamiltonian is in the matrix form. 
The numerical method for this specific problem is optimized in Ref.~\cite{Li:2021zaw}. In short, within each small time step $\delta x^+$, we treat $P^-_{KE}$ and $V_{\mathcal A}$ as time-constant and carry out matrix exponentiation in the momentum and coordinate space, respectively; the operation with $V_{qg}$ uses the fourth-order Runge-Kutta method in the momentum space. 

\section{Result}
In studying the phenomenon of jet momentum broadening inside a medium, we examine the expectation value of the transverse momentum square $\braket{p^2_\perp ( x^+)}$, and the related quenching parameter defined as $\hat q
  = \Delta\braket{p^2_\perp ( x^+)}/ \Delta x^+$.
\subsection{Eikonal analytical result}
In the eikonal limit of $p^+=\infty$, only the $V_{\mathcal A}$ term survives in the Hamiltonian, and the evolution operator reduces to the Wilson line, then $\braket{p^2_\perp ( x^+)}$ and $\hat q$ can be derived analytically using the Wilson line correlators.
Here, we present the derivation result for the single quark/gluon state, and a novel derivation for the quark-gluon state following Ref.~\cite{Li:2023jeh}.
\subsubsection{Single-particle state}
The Wilson line of a quark is
  $U_F(0,x^+; \vec x_\perp)\equiv\mathcal{T}_+\exp\bigg(
  -i g\int_{0}^{x^+}\diff z^+\mathcal{A}_a^-(\vec x_\perp, z^+)T^a
  \bigg)$,
in which $T^a$ is the SU(3) generator in the fundamental representation. Replacing $T^a$ by the generators in the adjoint representation, $t^a$, one gets the adjoint Wilson line for the gluon, $U_A(0,x^+; \vec x_\perp)$. 
The momentum transfer can be evaluated from the Wilson line correlator 
\begin{align}\label{eq:SF}
  \begin{split}
   S_F(0,x^+;r)
    = &\frac{1}{N_c}
    \Tr
    \expconfig{U_F^\dagger(0,x^+;\vec x_\perp) 
    U_F(0,x^+;\vec y_\perp)}
  =
    e^{
    -C_F
    g^4 \tilde{\mu}^2 x^+
    \left[
      L(0)
    -
    L(r)
    \right]
    }
    \;,
  \end{split}
\end{align}
with $r=|\vec x_\perp-\vec y_\perp|$ and $ L(r) 
=\int_{\p} e^{-i \vec p_\perp \cdot (\vec x_\perp-\vec y_\perp)}/(m_g^2+\vec p_\perp^2)^2
=m_g r K_1(m_g r)/(4\pi m_g^2)$,\footnote{Here and throughout the paper we use the shorthand notation $\int_{\p}\equiv \int \diff^2 p_\perp/(2\pi)^2 $ and $\int_{\r}\equiv \int\diff^2 r_\perp $.} as
\begin{align}\label{eq:psq_eik_res}
  \begin{split}
    \braket{p_\perp^2 (x^+)}_{Eik}
    =& \braket{p_\perp^2 (0)}
    -
    \nabla_r^2 S_F(0,x^+;r)|_{r=0}
    \;.
  \end{split}
\end{align}

The quenching parameter $\hat q$ follows as,
\begin{align}\label{eq:qhat_Eik_res}
  \begin{split}
  \hat q_{Eik}
    =
    -C_F
g^4 \tilde{\mu}^2 
 \nabla_r^2
  L(r)\big|_{r=0}
  =
  \frac{C_F
  g^4 \tilde{\mu}^2 }{4\pi}
  \Biggl\{
    \log\left[1+\frac{1}{(m_g a_\perp/\pi)^2}\right]
    -
    \frac{1}{1+(m_g a_\perp/\pi)^2}
  \Biggr\} 
    \;.
  \end{split}
\end{align}
In analogy, one gets the gluon $\hat q$ replacing $C_F$ by $C_A=N_c$ in Eq.~\eqref{eq:qhat_Eik_res}.
\subsubsection{Quark-gluon state}
The quark-gluon Wilson line is built as the tensor product of a quark and a gluon Wilson line,
$
    U_{qg}(0,x^+; \vec x_\perp, \vec y_\perp) \equiv U_F(0,x^+; \vec x_\perp)\otimes U_A(0,x^+; \vec y_\perp)
$.
The probability distribution of the quark-gluon state is then given by Wilson line correlator $ U^\dagger_{qg} U_{qg}$, and the state's transverse momentum square can be calculated accordingly.
It contains three terms,
 \begin{align}\label{eq:p_qg}
  \begin{split}
    \braket{p_\perp^2 (x^+)}_{qg,c;Eik}
    =&\braket{\vec p_{q,\perp}^2 (x^+)}_{Eik}
    +\braket{\vec p_{g,\perp}^2 (x^+)}_{Eik}
    +2\braket{\vec p_{q,\perp}(x^+)\cdot \vec p_{g,\perp} (x^+)}_{c;Eik}
    \;.
  \end{split}
\end{align}
The first two terms are the same as Eq.~\eqref{eq:psq_eik_res} with the corresponding Casimir, and the third term depends on the initial color configuration and the quark-gluon separation~\cite{Li:2023jeh},
\begin{align}\label{eq:pperp_qg_red}
  \begin{split}
    \braket{\vec p_{q,\perp}(x^+)\cdot \vec p_{g,\perp} (x^+)}_{c;Eik}
    =&    
    \begin{cases}
      0, & c=3\otimes 8\\
      -\frac{N_c \sqrt{2}}{2}f_{12}, & c=3\\
      -\frac{\sqrt{2}}{2}f_{12}, & c=\bar 6\\
      \frac{\sqrt{2}}{2}f_{12}, & c=15
    \end{cases}
    \;,
  \end{split}
\end{align}
in which $c$ represents the initial color configuration of the state, and 
\begin{align}\label{eq:s12}
  \begin{split}
    f_{12} = \int_{\v}f_{Rel} (\vec v_\perp)
    \nabla^2_v L(v)\frac{2\sqrt{2}}{4 N_c [L(0)-L(v)]}
    \left\{e^{ - g^4 \tilde{\mu}^2  N_c [L(0)-L(v)] x^+}-1\right\} \;.
  \end{split}
\end{align}
The quantity $f_{Rel} (\vec v_{\perp})$ is the distribution function of the quark-gluon relative coordinate $\vec v_{\perp}=\vec x_{q,\perp}-\vec x_{g,\perp}$, and it can be obtained by integrating the wavefunction square over the center-of-mass coordinate $\vec R_\perp=z\vec x_{q,\perp}+(1-z)\vec y_{g,\perp}$, 
$  
    f_{Rel} (\vec v_\perp)
    \equiv
    \int_{\R} 
    \left|\tilde \phi (\vec x_{q,\perp},\vec x_{g,\perp})\right|^2 
$.

\subsection{Non-eikonal numerical result}
We perform the simulations in the $\ket{q}+\ket{qg}$ space, with the initial state as a single quark of $\vec p_\perp=\vec 0_\perp$ at a finite $p^+$. 

\begin{figure*}[tbp!]
  \centering
\includegraphics[width=.85\textwidth]{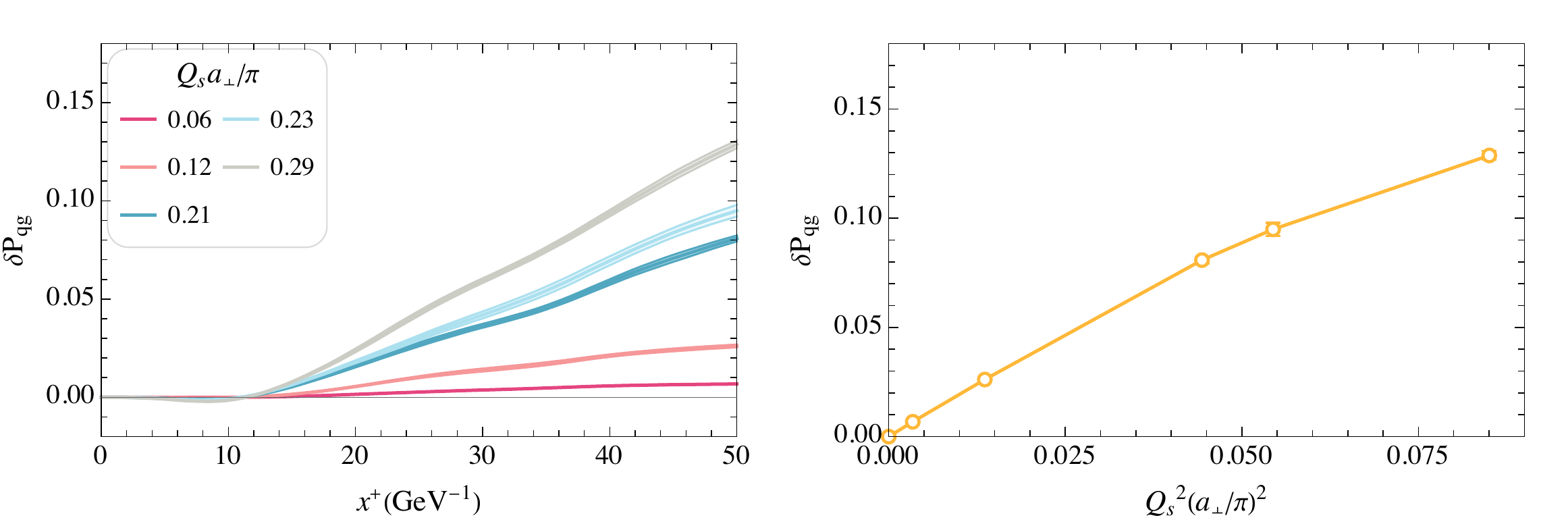}
  \caption{
  The medium-induced gluon emission. 
  Simulation parameters: $p^+=17~\GeV$, $N_\perp=16$, $L_\perp=50~\GeV^{-1}$, $L_\eta=50~\GeV^{-1}$ and $K=8.5$.
 }
  \label{fig:delta_Pqg}
\end{figure*}

To quantify the medium-induced gluon emission, we define $\delta P_{\ket{qg}}$ as the difference of the probability of the quark jet in the $\ket{qg}$ sector in the medium and that in the vacuum,
\begin{align}\label{eq:delta_Pqg}
  \delta P_{\ket{qg}}(Q_s, x^+)\equiv P_{\ket{qg}}(Q_s, x^+)-P_{\ket{qg}}(Q_s=0, x^+)\;.
\end{align}
The result is shown in Fig.~\ref{fig:delta_Pqg}.  
In the left panel, we see that the $\delta P_{\ket{qg}}$ curve forms a slight dip at an early time, then after around the point $x^+=12~\GeV^{-1}$, grows linearly in time.
Additionally, a larger $\ket{qg}$ component develops as $Q_s$ increases.
The right panel shows that the $\delta P_{\ket{qg}}$ of the final state ($x^+=L_\eta$) is approximately proportional to $Q_s^2$ (c.f.~\cite{Zhang:2021tcc} has a similar observation).

\begin{figure*}[tbp!]
  \centering
\includegraphics[width=.85\textwidth]{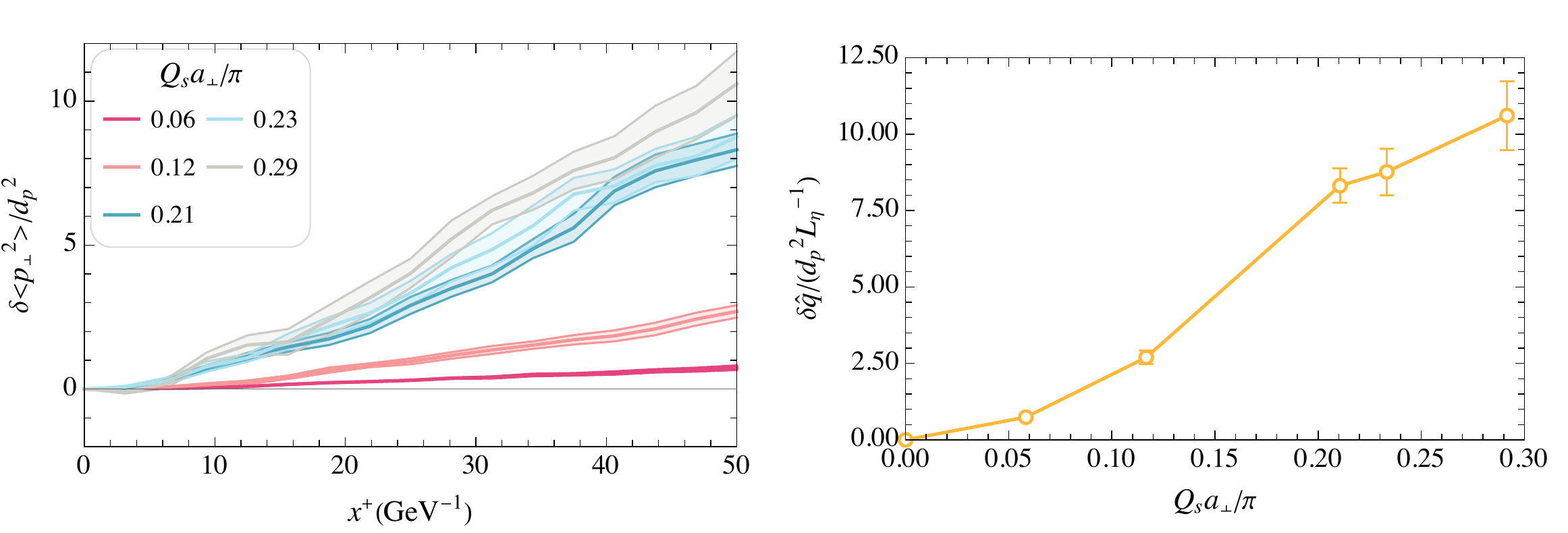}
  \caption{
  The non-eikonal correction to $ \braket{p_\perp^2}$ and $\hat q$. Simulation parameters are the same as in Fig.~\ref{fig:delta_Pqg}.
 }
  \label{fig:delta_qhat}
\end{figure*}

We then analyze the non-eikonal and radiative correction to the momentum broadening. We define $\delta \braket{p_\perp^2}$ and $\delta \hat q$ as the difference of the quantity that is calculated from the total momentum of the quark jet in the $\ket{q}+\ket{qg}$ space, and the eikonal result of a bare quark [as in Eq.~\eqref{eq:qhat_Eik_res}],
\begin{align}\label{eq:delta_qhat}
  \delta \braket{p_\perp^2}\equiv \braket{p_\perp^2}-\braket{p_\perp^2}_{Eik}, \qquad \delta \hat q \equiv \hat q-\hat q_{Eik}\;.
\end{align}
The results are shown in Fig.~\ref{fig:delta_qhat}: $\delta \braket{p_\perp^2}$ increases over the evolution time at various $Q_s$, and $\delta \hat q$ extracted from the final state $\delta \braket{p_\perp^2}$ increases non-trivially when $Q_s$ increases.

\section{Summary}
We present a study on the momentum broadening of in-medium jet evolution using the tBLFQ approach \cite{Li:2021zaw,Li:2023jeh}, a non-perturbative light-front Hamiltonian formalism.
We first provide a novel analytical derivation of the eikonal expectation value of the quark-gluon state's transverse momentum for any color and spatial distribution.
We then perform the numerical simulation of the real-time jet evolution in the Fock space of $\ket{q}+\ket{qg}$ at various medium densities.
With the obtained jet light-front wavefunction, we extract the gluon emission rate and the quenching parameter.
We find their non-eikonal contributions sizable, time-dependent, and associated with saturation scale. 

\acknowledgments
We are very grateful to Guillaume Beuf, Fabio Dominguez, Miguel A. Escobedo, Xabier Feal, Sigtryggur Hauksson, Cyrille Marquet, Wenyang Qian, Andrecia Ramnath, Andrey Sadofyev, Konrad Tywoniuk, James P. Vary, Xin-Nian Wang, and Bin Wu for helpful and valuable discussions. 

XB is supported by new faculty startup funding by the Institute of Modern Physics, Chinese Academy of Sciences, by Key Research Program of Frontier Sciences, Chinese Academy of Sciences, Grant No. ZDBS-LY-7020, by the Natural Science Foundation of Gansu Province, China, Grant No. 20JR10RA067, by the Foundation for Key Talents of Gansu Province, by the Central Funds Guiding the Local Science and Technology Development of Gansu Province, Grant No. 22ZY1QA006 and by the Strategic Priority Research Program of the Chinese Academy of Sciences, Grant No. XDB34000000.
ML and CS are supported by Xunta de Galicia (Centro singular de Investigacion de Galicia accreditation 2019-2022), European Union ERDF, the “Maria de Maeztu” Units of Excellence program under project CEX2020-001035-M, the Spanish Research State Agency under project PID2020-119632GB-I00, and European Research Council under project ERC-2018-ADG-835105 YoctoLHC.
TL is supported by the Academy of Finland, the Centre of Excellence in Quark Matter (project 346324)
and project 321840.
This work was also supported under the European Union’s
Horizon 2020 research and innovation by the STRONG-2020 project (grant agreement No. 824093). The content of this article does not reflect the official opinion of the European Union and responsibility for the information and views expressed therein lies entirely with the authors.


\end{document}